\documentclass{article}

\usepackage{PRIMEarxiv}

\usepackage[utf8]{inputenc} 
\usepackage[T1]{fontenc}    
\usepackage{hyperref}       
\usepackage{url}            
\usepackage{booktabs}       
\usepackage{amsfonts}       
\usepackage{nicefrac}       
\usepackage{microtype}      
\usepackage{lipsum}
\usepackage{fancyhdr}       
\usepackage{graphicx}       
\graphicspath{{media/}}     

\title{Understanding Digital Government Transformation}

\author{
  Mamdouh Alenezi \\
  College of Computer and Information Sciences \\
  Prince Sultan University \\
  Riyadh, Saudi Arabia\\
  \texttt{malenezi@psu.edu.sa} \\
}

\begin{document}
\maketitle

\begin{abstract}
In today’s era of innovation of technological progression, digitalisation has not only transformed individual lives but also has a prominent influence on business activities. The world is surviving in a global yet complex technological progression that not only changes the lives of civilians but is also transforming the public, private, and academic spheres of life. This research focuses on the digitalisation of governments, their challenges, and success factors. It is found that government faces difficulties in formulating strategies, proper planning, execution strategies, and a lack of organised information and expertise. However, success can be achieved by working on capabilities of the future workforce, creating leaders for tomorrow, generating digitalisation capabilities, and bringing a purpose-driven digitalisation before digital government transformation. Overall, the study’s findings suggest that digital government transformation creates value, enhances relations, improves service delivery, grows economy, pushes economic activities, enhances citizen engagement, increases the policy implementation and their efficiency, and affects business growth positively.
\end{abstract}


\section{Introduction}

The world today is striving towards a global yet complex technological progression that has not only changed the lives of civilians but is also transforming the public, private, and academic spheres of life. In addition, this technology-driven change extends opportunity for nations to converge advancements with an aim to create a more human-centric and inclusive future \cite{abdulrahim2020covid}. It is very important to understand the reason prior to discussing the significance of digital transformation. The rationale to shift towards digital transformation is linked with changing needs of consumers and market dynamics, existing threats with the traditional and conventional methods, and need of innovation to sustain the competitiveness \cite{morakanyane2020determining}. One reason could be to leverage emerging technologies. With the emergence and prioritization of the digital economies facilitating a user-centric and seamless experiences, citizens now anticipate public sector to be highly accessible, flexible, and efficient. Governments across the globe are transitioning towards digital transformation (DT) as a strategic motive to enhance the service performance, improve consumer experience, streamline the activities and operations, and develop new business models \cite{mergel2019defining}. This paper aims to provide an overview for the government institutions regarding the impact of digital transformation through the analysis of statistics and information, highlighting the reliance on digitisation.

The adoption of DT in governmental institutions is continuously changing the citizen’s expectations \cite{mergel2019defining}. E-government is the latest incorporation of digital government transformation. E-government is a digital application transformation that facilitates country services and products for its industry and civilians by the implementation of information technologies and electronic tools \cite{amayah2013determinants}. These expectations include real-time and high-quality digital services. Similarly, European Commission \cite{eurcomm} stated that the objective of digital transformation in the public sector tends to focus on creating new forms of relationships, building new frameworks of services delivery, and generating innovative ways of working with stakeholders. The governments need to address such decisions very critically. Furthermore, the strong and fast adoption of digital government transformation can support governments heavily while working on these objectives. Similarly, the demands of politicians, businesses, and civilians tend to be extremely important as they determine the efficiency of governments. Technological changes are heavily experienced by these individuals in their work, life, and environment \cite{jonathan2021public}. Hence, one of the aims of this study is to identify how governments can address these expectations. Specifically, the paper will evaluate the digital government transformation. In addition, the study will evaluate the influence of digital transformation on the decision making of governments, policymakers, and researchers by analysing the success factors and challenges of the DT.

\section{Digital Transformation}

Digital Transformation refers to the leveraging of business capabilities and digital technologies for the sake of enhancing customer experience, operational processes, and business models which create customer value \cite{morakanyane2017conc}. In the current era, organisations adopt emerging technologies to achieve rapid development, which in return changes their business competition positively whereby, such organisations create new ways of operations \cite{loebbecke2015reflections}. Moreover, according to Fitzgerald \cite{fitzgerald2013starbucks}, all the organisations and across the industries are affected by digital transformation; such effect pushes organisations to adopt digital transformation to remain competitive in their market or segment. 

Precisely, digital transformation enhances the value chain, stakeholders’ relationships, and efficient deliveries \cite{ebert2018digital}. Moreover, digital transformation provides an edge to enterprises by increasing the ease of monitoring, controlling, and executing the strategies while reducing barriers of connectivity and burdens of responsibilities. Furthermore, DT is regarded as the backbone of sustainability, survivability, and competitive edge of the business through efficient utilisation, innovation, and operational integration \cite{fletcher2020digital}. In simple words, all the activities, operations, policy execution are synchronised by interconnectivity which gives access to each other. As a result, businesses can effectively monitor, control, and design strategies accordingly. Hence, it has now become the need of time for the organisations to integrate digitalisation to sustain its competitive edge.

\section{Elements of Digital Transformation}

The term digital transformation, which is mostly used in the private industrial segment, forces the organisations to adopt new and advanced technologies so remain differentiated in the world of the internet due to the provision of services offline and online \cite{berman2012digital}. The elements of digital transformation include transforming value creation, relations with citizens and organisational culture, and service delivery. 

\subsection{Transforming to Create Value}

The value chain could lead to a positive shift by acquiring digital transformation. Berman \cite{berman2012digital} determined that the activities such as co-funding, co-distribution, co-marketing, co-production, co-creation, and co-design are heavily improved by the incorporation of digital transformation. This is because it increases the connectivity and reliability between these departments, which then result in higher performance \cite{nogravsek2014government}. Thus, integration of DT leads to overall efficiency in the operations through higher interconnectivity that adds value to the firm.

\subsection{Transforming to Enhance Relations}

Digital transformation is considered a technological revolution that increases the consumer’s expectations, whereby businesses can fulfil them effectively by the use of digital transformations. This indicates that DT can lead to the accomplishment of high expectations, that is regarded as a cultural modification. Similarly, government businesses inject consumer engagement, integrity, and transparency by implementing digital solutions.

\subsection{Transforming to Enhance Service Delivery}

The digital transformation provides new ways of service delivery, customer interactions, and delivery modes \cite{berman2012digital}. It supports the value chain actors, including connections between consumers, external producers, and suppliers. These benefits pinch enterprises to adopt digital transformation for the sake of efficiency. Hence, it can be said that digital transformation is basically upgrading the value chain to achieve the enterprise’s objectives. 

These benefits can also be utilised in government institutions by adopting digital transformation. It is now clear that there is a huge need for digital government transformations. This study emphasises transformational government, digital government, e-governance, and e-government.

\section{Digital Government Transformation}

\subsection{E-government and Technology}

Technology is an essential element of public infrastructure. It supports creating value in the society at large, private, and public sector through sustainable innovation. In addition, it is adopted to treat economic disruptions, environmental disruptions, and transferable diseases \cite{nielsen2016potential}. However, if the technology is used in the public sector, different things are at stake that includes trust in institutions and government, risk adversity, inter-governmental models, governance, legal traditions, organisation and operations of government, socio-economic factors, culture, and technology itself \cite{meyerhoff2019governance}. 

In the last two decades, researchers have intensively studied the digital government transformation giving many definitions of e-government. Rooks, Matzat, and Sadowski \cite{rooks2017empirical} defined that e-government uses Information communication technologies (ICTs) to increase facilities for its residents along with providing government information to its residents by using information communication technologies and internet. The literature on this topic considers transformational government, digital government, and e-government similar to digital government transformation.

In simple words, these concepts are interrelated, sharing a similar nature of the ground \cite{mergel2019defining}. In addition, they evaluate the impact of using information communication technologies on culture, organisational process, and service delivery along with the influence on value creation. At first, it is necessary for the government to break the traditional process to bring transformation in the operations in the form of digitalisation. Primarily, it includes the re-engineering of process and structure known as Business Process Engineering (BPR) \cite{weerakkody2021resurgence}. Therefore, incorporation of ICT facilitates in seamless and efficient government operations to improve citizen experience and satisfaction.

The main idea of incorporating electronic government is to improve and facilitate governmental services with the use of communication and information technology. In simple words, the traditional governments are transformed into e-governments to integrate the administration and enhance connectivity among the policymakers \cite{iskender2015analysis}. Although the idea looks very simple, yet its installation and implementation tend to be very complex \cite{ma2018does}. The complexity is related to the availability of various factors, including economic, legal, political, organisational, social, and technical. Hence, the success of transforming becomes subjective to how efficient transformation is carried out. 

\subsection{E-government Characteristics}

The comprehensive use of technology has the ability to offer several social and economic benefits. The digital transformation enables inclusion allowing small and large organisations to sell their product and service offerings online while competing on a global scale. Similarly, e-government facilitates country services and products for its industry and civilians by the implementation of information technologies and electronic tools \cite{pedersen2018government}. Moreover, digitalisation contributes to increasing efficiency. Digital technologies allow companies as well as government entities to make better use of resources, labour while leveraging the intelligence, speed, and reliability of digital tools. According to Amayah \cite{amayah2013determinants}, DT refers to organisational changes which use new business models and new digital technologies to enhance consumer experience and organisational performance.

It is essential to digitise public services as it helps in economic growth, pushes economic activities, enhances citizen engagement, and positively affects business growth. In the last decade, the use of digital transformation has trended in developed as well as the developing countries \cite{alvarenga2020digital}. Most importantly, there is increasing problem of unskilled personnel in governmental organisations and a lack of knowledge in the staff \cite{alvarenga2020digital}. Weerakkody et al. \cite{weerakkody2016digitally} explained that these problems could be solved as the adoption of digital transformation leads to innovative solutions in political, social, and economic areas whereby the strong decision-making process is followed. In addition, these changes are necessary to be understood by all the individuals who prepare and implement policies or make a decision regarding government practices \cite{lee2018proposing}. These individuals include researchers, government executives, and policymakers. Hence, considering these benefits, it is very crucial for the government to adopt DT for the purpose of predicting and understanding the ongoing changes.

Digital Government Transformation can support the Sustainable Development Goals (SDGs) in multiple directions. The development of digital inclusion in the government administration supports the fast and efficient execution of operations. According to Nielsen \cite{nielsen2016potential}, digital government transformation reduces the administration burden and increases productivity, acquires higher productivity, and facilitate in the adherence to Sustainable Development Goals (SDGs). Moreover, effective use of technology with efficient re-engineering and organisational processes and regulations also bring employment, sustainable economic growth, and inclusive and transparent society. Thus, DT in the government promotes sustainable measures for the betterment of the environment. 

\section{Digital Government Transformation Challenges}

With the emerging era of engaging the public sectors with digital transformation, this journey combats a number of challenges. To discuss most of the prominent ones, the digital government transformation challenges can be generally categorised into two common categories; internal and external challenges. According to Lindgren and van Veenstra \cite{lindgren2018digital}, internal challenges refer to the difficulties that are raised within the public sector, whereas the external challenges are the ones that arise from the bodies that are not the immediate organs of the government but hold external effects on it.

As to identify the internal challenges, studies reveal that to attain a goal or objective, an organisation must have a mutual vision and mission for it \cite{shevtsova2020conceptual}. Similarly, the initial issue a government organisation faces when it comes to digital transformation is the management of formulating an appropriate digital transformation strategy. The public sectors must be aware of their digital/IT needs, so an appropriate strategy must be designed. Research by Jonathan et al. \cite{jonathan2021public} described that none of the interviewees – who were involved in the research – were aware of any kind of digitalising strategy. Some of the interviewees had heard about it, but they find it problematic. Also, proper planning for sustaining the strategy is clearly associated with this challenge. 

Organisational structure and its culture are also considered a significant challenge that fosters the digitalisation transformation process in organisations. Public sectors are the most vulnerable organisations that commonly face these challenges \cite{pedersen2018government}. The employees themselves show a lack of motivation to adopt the digital changes and willingness towards continuing the existing system. According to studies that conducted interviews for analysing the prominent challenges, most of the respondents were found to be happier and satisfied with the existing typical analogue culture they use in their organisation \cite{jonathan2021public}. Therefore, in the way to bring up a digitalised transformation in a public sector, organisational culture becomes resistant. 

Last but not least, information security is one of the radical challenges among all. According to Mamonov and Benbunan-Fich \cite{mamonov2018impact}, digital services can never be categorised as services to the public if the sensitive information in them is not secure. When anything is brought up to a digital platform, the level of information security threats gets increased. Both risks for information security threats as well as the lack of organised information secured tactics go in parallel to each other. Digital tactics for information security is highly necessary for any public sector; otherwise, the system will no longer be beneficial in terms of providing services to the public.

In terms of external challenges, there exist two major challenges that are responsible for fostering the emergence of digital growth in the public sector. The first is lack of expertise in the market, and the other is stakeholder relationships. Evidence suggested that the availability of skilled and trained experts who have solid grips on digital technologies is considered to be the foremost requirement to digitally transform any sector, whether a private or public sector firm \cite{gil2018digital}. It is also evident that skilled technical individuals can easily be found, but the main problem lies within the lack of people who have firm experience in recent forms of digitalisation and training the relevant staff accordingly \cite{jonathan2021public}. One of the most affected countries who possess lack of trained and expertise in the market are the developing countries. Nevertheless, strict public sector regulations and protocols also affect the expert individuals to get attracted towards the public sector digitalisation processes. 

On the other hand, Nambisan, Wright, and Feldman \cite{nambisan2019digital} explained that the stakeholder relationship can also play a challenging role for the public sectors to incorporate the digital transformation if not maintained accordingly. Primarily, these stakeholder relationships are associated with the private organisations, but they can be associated with the government departments as well, depending on the interests of the public sector organisations. Even though the main objective of digitally transforming the public sector is to enhance transparency and accountability, but multiple engaging stakeholders who can assist in different dimensions of decision making is one of the significant purposes \cite{jonathan2021public}. However, the management of the stakeholder relationship, including their participation, is a potential challenge for public sector organisations.

\section{Digital Government Transformation Success Factors}

Organisations certainly experience challenges and hurdles to incorporate digital transformation, but implementing appropriate strategies and including some of the prominent success factors may help gain the objective. Evidence suggested that one of the prior success factors is to enhance the capabilities of the future workforce \cite{kattel2019estonia}. Success can only be achieved if an effective environment is provided to the employees where they can work and represent their talents as well as achieve new capabilities. With the initiation of digital transformation, it is immensely important to establish digital capabilities for the working and future workforce. Once the employees get trained over the digital transformation, they start to work as prime integrators of the system, where the integrators refer to the ones who adopt the latest changes of the digital frameworks and integrate them into the existing organisational culture. 

The second success factor is to create leaders who are digitally well-aware and demonstrated. According to a research survey conducted by Liferay \cite{liferay2017}, it has been examined that more than 30\% of the respondents involved in the survey used to believe that organisational change was the biggest barrier towards digitalising their public sector. The challenging aspect is to break and transform the traditional analogue system of the organisation and bring a different way for the public sector. Therefore, it is not only important to bring transformation across the board, but it is also potentially needed to create a team of leaders who play a key role in the success of digital transformation \cite{pedersen2018government}. It is because a digital transformation in the public sector can only be seen if the leaders are strongly willing to implement it. In this regard, they must be aware of all the pros and cons of the digitalisation process, as well as the reason behind integrating it into the public sector organisations. 

Another significant success factor is related to the intercommunication between the digital and non-digital approaches. When a public sector incorporates digital transformation, it is crucially important to formally communicate the actual narrative for bringing up the change. According to Morakanyane et al. \cite{morakanyane2020determining}, the doer must help understand the workers comprehensively why this digitalising is being transformed. This helps in building a concept for the employees and also creates a passion for working for them. The narrative to demonstrate the reasons behind digital transformation should be rich, covering every aspect of the latest changes to excite the employees to work accordingly. This would also cover describing the short and long term goals as well as the key performance indicators (KPIs).

Another factor that determines the success of digital transformation in the public sector is to enable accessibility to everyone in the country \cite{morakanyane2020determining}. Different citizens may have different needs, but their expectations from the government remain the same. The transformation must ensure that the public sectors are easily accessible to every citizen by using any device. Currently, only 76 out of 193 United Nations (UN) member states are completely accommodating web access nationwide, which is relatively very low. Therefore, it is high time for the government to change the typical procedure of engagement and enable everyone to participate digitally in multiple domains of the public sector.

A far better success factor is to propose a purpose-driven and seamless experience into the digital transformation of public sectors. It is predefined that the target audience might have different needs and behaviours depending on the nature of service they require from the government. Therefore, incorporating a purpose-driven digital transformation into the government agencies can bring positive outcomes in terms of successful implementation of the transformation. Every government agent must have a relevant platform that is desired by the audience. For example, audiences applying for a visa to one country is different from the audience applying for a vehicle license. So, there is a need to integrate certain relevant and appropriate digital corners in the transformation that assist the incoming audience to their best services. It is evident because the audience who experience satisfactorily through any website is more than 105\% likely to use such websites as their primary source to have services \cite{kattel2019estonia}. A more proportion of this audience is also prevalent to recommend these sites to friends and their community, which further enhances the application and services of the digital transformation.

\section{Discussion}

Digital transformation in the government or public sector refers to different and innovative ways of engaging and working with stakeholders, developing frameworks for efficient service delivery mechanisms and forming new relationships. Nevertheless, ahead of the availability of reports of leading corporations, i.e. Deloitte, there is limited systematic empirical evidence found regarding the way public administrations define digital transformation in their daily practices, what the anticipated outcomes are, and how they approach digital transformation projects. The terms such as digital transformation, digitalisation or digitisation are used interchangeably in the studies \cite{mergel2019defining}. Considering the role of digital transformation in government, studies have observed that digital transformation are the changes in service mode delivery, and are also new types of direct engagement with clients, for instance, via social media sites to adapt services and products as per changing consumer demands or needs \cite{mergel2019defining}. The conception of e-government over the past two decades has been studied extensively, and researchers like Rooks, Matzat and Sadowski \cite{rooks2017empirical} differentiated narrow and broad explanations of e-governments. In terms of narrow definition, the emphasis is on the use of information communication technologies (ICTs) to deliver services to people, while broad definition includes the use of ICTs and the internet to provide government and administrative information to the people. Other definitions, as per Ma and Zheng \cite{ma2018does}, of digital transformation focus on the engagement with people via ICTs. Other than the definitions, the role or the effectiveness of e-government on companies is still a disputed issue. Nevertheless, the advantage of e-government emphasises largely on the improvement of service delivery, leading to increase efficiency in government performance.

Digitalisation has been adopted in developed countries, and Denmark is considered one of the leading countries that have established e-government programs. As per the UN e-government survey, Denmark was rated to be one of the mature countries regarding e-government \cite{meyerhoff2019governance}. The collaborative, consensus-seeking, and coordinated approach that has taken to the digital transformation of the Danish government sector are a few of the key elements behind the success of Danish authorities in e-government. Additionally, the formal and informal discussions are the norm of e-government strategies in Denmark, and it is unluckily the end-user group, academia, and private sector that are not part of the governance model, as they could assist in ensuring a holistic approach to the use of technology in government sector.

E-government strategies in Denmark are facilitated by strategic initiatives associated with the necessary business case model and IT-project, which assist in reducing the risk of failure and further support active benefit of understanding at both strategy and project levels \cite{meyerhoff2019governance}. Similar findings have been observed in the study of Janowski \cite{janowski2015digital} who argued that the concept of digital government has become widespread, and the evolution model of digital government can now be categorised into technology or digitisation in government, electronic or engagement governance, policy-driven or contextualisation electronic governance and electronic or transformation government. Gong, Yang, and Shi \cite{gong2020towards} discussed the understanding of digital transformation in government by asserting that it is a slow process with adaptations in numerous structural elements, affecting the entire administrative system from local to the federal level, including both incremental and radical alterations. Flexibility increases together with the development of digital transformation. However, the flexibility more relies on bureaucratic levels and organisational elements. 

Other than the adoption of digitalisation in the public sector, there are certain challenges, which are faced when implementing it on a wider level. For instance, the study of Jonathan et al. \cite{jonathan2021public} pointed out the management of developing a relevant and suitable digital transformation strategy is one of the key issues faced by the government at initial levels. The public sector should be conscious of their IT/digital requirements, and hence appropriate strategy should be crafted accordingly. Additionally, proper planning and effective mechanism are important for sustaining the strategy and addressing challenges. The culture and structure of the organisation, as per Jonathan et al. \cite{jonathan2021public}, is a considerable challenge that increases the digital transformation procedure in the organisation. The vulnerability is high among public sector organisations as they commonly encounter such challenges. The workforce themselves exhibit low motivation and enthusiasm to adopt digital alteration and show eagerness to continue with the existing system.

Considering the external problems, there are two main challenges that restrict the increase of digital growth in the governmental sector. Firstly, lack of experience and stakeholder relationship are the two challenges found in the public sector, which affects the growth of digitalisation. Evidence from the research of Gil-Garcia et al. \cite{gil2018digital} showed that the availability of trained and skilled experts who possess better understanding and knowledge regarding digital technologies are viewed to be the primary need for digital transformation in both the public and private sector. It is also noted that technically skilled people are available, yet the issue is with people who lack experience with reference to recent digitalisation and training the workforce accordingly \cite{jonathan2021public}. The developing countries are viewed to be highly affected in this regard, besides strict protocols, and public sector regulations are other challenges that affect public sector organisations to attract experts in increasing their digitisation process.

In contrary to the issues, there exist success factors as well, which can be accomplished if employees are provided with an effective environment where they can represent and work their talents and attain new capabilities. It is highly important with the start of digital transformation to create digital capabilities for the working and for the future workforce. The digital training of employees is important because they will be the key integrators of the tech-based system, where they will have the key role in integrating digital framework and developing a new culture by replacing or reshaping the existing one. Secondly, the role of leadership is also important; who must be knowledgeable and have experience in addressing digital changes. Morakanyane et al. \cite{morakanyane2020determining} add that accessibility in the public sector is the factor that determines the successful integration of digital transformation. A different citizen may have a varying requirement, and therefore their expectation from the authorities remains the same. The digital transformation should be developed according to the public needs and needs to be accessible for a citizen with multiple different options. Therefore, it is pivotal for the government to adapt to changes and work on the environment that facilitates technological changes in the organisation. 

\section{Conclusion}

This research emphasised digital transformation in government and examined how the digital transformation over the past few decades and have transformed and what challenges/issues the public sector faces in integrating such technologies. It is observed that Denmark is one of the leading developed countries to adopt digital transformation due to effective e-government strategies and a suitable environment. However, developing countries are found to be lacking in adopting digital transformation due to the lack of tech culture in their public and private organisation. Additionally, the lack of trained and skilled employees is also one of the key challenges that restrict digitisation in the government sector. The research highlights the effective and favourable environment, skilled workforce, leadership, and governmental policies and regulations are the key success factor that may encourage and facilitate the quick adaption of digital transformation in public sector organisations.

\bibliographystyle{unsrt}  
\bibliography{references}

\end{document}